\documentclass{wspr9x6}

\usepackage{wasysym}

\begin{document}

\title{SYMMETRY ENERGY}

\author{P.~DANIELEWICZ$^*$}

\address{National Superconducting Cyclotron Laboratory, Michigan State University,\\
East Lansing, Michigan 48824, USA\\
$^*$E-mail: danielewicz@nscl.msu.edu}

\begin{abstract}
Examination of symmetry energy is carried out on the basis of an elementary
binding-energy formula.  Constraints are obtained on the energy value at
the normal nuclear density and on the density dependence of the energy at subnormal
densities.
\end{abstract}

\keywords{symmetry energy, mass formula, nuclear matter, neutron matter}

\bodymatter

\section{Introduction}\label{sec:Intro}

In nuclear physics, the symmetry energy is first encountered within the elementary
Bethe-Weizs\"{a}cker formula for nuclear energy:
\begin{equation}
\label{eq:BW}
{ E = - a_V \, A + a_S \, A^{2/3}
+ a_C \, \frac{Z^2}{A^{1/3}} +} { a_A \, \frac{(N
-  Z)^2}{A} } { + \Delta} \, ;
\end{equation}
it is the change in nuclear energy associated with changing neutron-proton asymmetry
$(N-Z)/A$.  In nuclear matter, the energy per nucleon, dependent on neutron $\rho_n$ and proton $\rho_p$
densities, may be represented as a sum of the
energy $E_0$ for symmetric nuclear matter and the correction $E_1$ associated with the asymmetry,
\begin{equation}
  E(\rho_n, \rho_p) = E_0 (\rho) + E_1 (\rho_n, \rho_p) \, ,
\end{equation}
where $\rho=\rho_n +\rho_p$.
The charge symmetry of nuclear interactions, which is the symmetry under the interchange of neutrons and protons,
requires that the correction be quadratic in the asymmetry, for small asymmetries:
\begin{equation}
  E_1 = E - E_0 \simeq { S(\rho)} \left( \frac{\rho_n-\rho_p}{\rho} \right)^2 \, .
\end{equation}
Microscopic calculations, such as \cite{bom91}, indicate that the quadratic approximation yields
a very good representation for the energy in nuclear matter, up to the limit of neutron matter with
asymmetry of~1, over a wide range of net densities.  As a consequence, the energy in nuclear matter
over a broad range of parameters can be described exclusively in terms of the energy of
symmetric matter $E_0(\rho)$ and the symmetry-energy coefficient $S(\rho)$.

Much effort has been dedicated in the past to the understanding of $E_0(\rho)$ and less so to
$S(\rho)$ whose features remain more obscure \cite{bro00}.  The energy $E_0(\rho)$ minimizes at the
normal density $\rho_0$, reaching there the value of $E_0 \approx - 16.0$~MeV.  The uncertainties in $S$ hamper
predictions for neutron stars whose structure depends on pressure
in neutron matter \cite{lat01}.
In pure neutron matter, the energy is $E=E_0+S$ and the pressure is
${ P = \rho^2 \, dE/d\rho \simeq \rho^2 \, dS/d\rho}$ close to $\rho_0$, as $E_0$ minimizes at $\rho_0$.
In the calculations of neutron-star structure \cite{lat01}, a correlation is found,
${R \, P^{-1/4} \approx \mbox{const}}$, between the radius $R$ of a neutron star of a given mass and
the pressure $P$ in neutron matter at a given density $\rho \sim \rho_0$.

\section{Binding Formula}

When examining the standard energy formula (\ref{eq:BW}), we notice that the symmetry energy
has a volume character: it changes as $A$ when the neutron and proton numbers are scaled
by one factor.  The formula lacks a surface symmetry term that would change as $A^{2/3}$ with the
change in nucleon number.  A question to ask is whether there should be such a term.
Let us look at the surface energy.  We can write it as $E_S = a_S \, A^{2/3} =
\frac{a_S}{4\pi \, r_0^2} \, 4\pi \,
r_0^2 \, A^{2/3}
=
\frac{a_S}{4\pi \, r_0^2} \, {\mathcal S}$.  The ratio
$\frac{E_S}{\mathcal S} = \sigma = \frac{a_S}{4\pi \, r_0^2}$ is surface tension, the work that needs
to be done per unit area when changing the surface area of the nucleus,
such as in deforming the nucleus.  The work needs to be done to compensate lost binding, as nucleons
close to the surface are less bound, due to fewer neighbors, than in the interior.  The formula (\ref{eq:BW})
states that in the interior the nucleons are less bound in a more asymmetric nucleus.  In that case,
the energetic price for increasing the surface should drop, i.e.\
$\sigma = \frac{\partial E_S}{\partial {\mathcal S}}$
should decrease with asymmetry, in the more general definition of tension.

\begin{figure}
\centerline{\includegraphics[width=.52\linewidth]{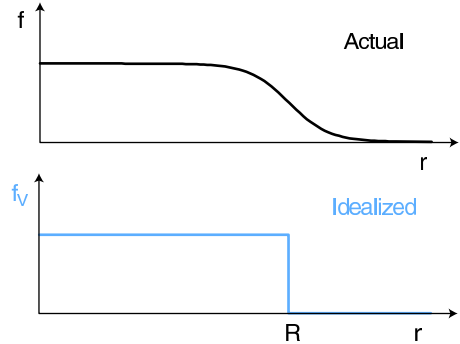}}
\caption{\label{fig:gibbs}
Gibbs' \cite{gib48} construction for defining volume and surface contributions.}
\end{figure}

\begin{figure}
\centerline{\includegraphics[width=.52\linewidth]{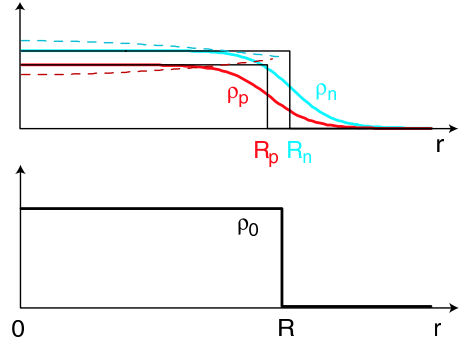}}
\caption{\label{fig:gibbs2}
In a two-component system, the surfaces for the two components will be, generally, displaced
relative to each other.}
\end{figure}

As intensive, $\sigma$ should be expressed in terms an intensive quantity associated with the asymmetry,
which is the asymmetry chemical potential:
\begin{equation}
\mu_A = \frac{\partial \, E}{\partial \, (N-Z)} = \frac{1}{2} \left( \mu_n - \mu_p \right)
\, .
\end{equation}
To lowest order, under charge symmetry, the tension needs to be quadratic in $\mu_A$,
\begin{equation}
\label{eq:sigma}
  \sigma = \sigma_0 - \gamma \, \mu_A^2 \, .
\end{equation}
If the tension depends on asymmetry, so must surface energy.  On examining the
function $\Phi=\mu_A(N-Z)-E$, with the derivative $\partial \Phi/\partial \mu_A=N-Z$, the last
dependence is seen to produce the following
apparent paradox: some of the nuclear asymmetry $N-Z$ must be associated with the surface and not
the interior.  To answer the question on how particles can be attributed to the surface, one needs
to adapt a systematic approach the separation of quantities into volume and surface contributions.
Gibbs \cite{gib48} proposed to consider two copies of the system, actual
and an idealized reference copy where the interior densities of different quantities extend up to the
surface position, cf.\ Fig.~\ref{fig:gibbs}.
The idealized system represents one with only volume contributions to different quantities, while
the difference between the systems can be associated with the surface, $F_S = F - F_V$.  The
separation, however, depends on the position of the surface which must be set utilizing some auxiliary
condition.  For nuclei, it is natural to demand a vanishing surface nucleon surface number $A_S=0$, i.e.\
set the surface position at the sharp-edged sphere radius~$R$.  Since nuclei, though,
are binary systems, the surface positions that might be separately attributed
to neutrons and protons will be,
generally, displaced relative to each other, cf. \ Fig.~\ref{fig:gibbs2}.
In consequence, in spite of a vanishing surface nucleon number, $A_S = N_S + Z_S = 0$, we may have a
finite surface asymmetry $N_S-Z_S \ne 0$.

Having resolved the apparent paradox, from (\ref{eq:sigma}) we find
\begin{equation}\label{eq:ES}
  E_S  = \sigma_0 \, {\cal S} + \gamma \, \mu_A^2 \, {\cal S} =
E_S^0 + \frac{1}{4 \gamma} \, \frac{(N_S - Z_S)^2}{\cal S}
 =  E_S^0 + a_A^S \, \frac{(N_S - Z_S)^2}{A^{2/3}} \, .
\end{equation}
Here, the index 0 refers to symmetric matter and we have introduced a surface symmetry
coefficient $a_A^S$, with dimension of energy.  For the volume energy, within the standard formula,
we have
\begin{equation}\label{eq:EV}
  E_V = E_V^0 + a_A^V \, \frac{(N_V - Z_V)^2}{A}  \, ,
\end{equation}
where $a_A^V$ is the volume symmetry energy coefficient.  The Coulomb energy is temporarily
ignored.  The net energy and asymmetry are, respectively, $E=E_S+E_V$ and $N-Z=N_S-Z_S+N_V-Z_V$.  In the ground
state, the asymmetry should partition itself into the surface and volume contributions, in such a way as to
minimize the energy.  The result of the energy minimization can be, actually, written right away once one
notices that the surface and volume energies, quadratic in asymmetry, are analogous to the energies of
capacitors quadratic in the charge.  The energy of the coupled capacitors is quadratic in the net charge, with the square
divided by the net capacitance, yielding:
\begin{equation}\label{eq:Enet}
  E=E^0 + \frac{q^2}{2C}=E^0 + \frac{(N-Z)^2} {\frac{A}{a_A^V} + \frac{A^{2/3}}{a_A^S}} \, .
\end{equation}

Adding now the Coulomb term, we arrive at the modified energy formula
\begin{equation}\label{eq:BWmod}
  E  =  - a_V \, A + a_S \, A^{2/3}
+ a_C \, \frac{Z^2}{A^{1/3}}  +  \frac{a_A^V}{1+ A^{-1/3} \, {a_A^V}/{a_A^S}
} \, \frac{(N
-  Z)^2}{A} \, .
\end{equation}
Compared to the standard formula (\ref{eq:BW}), the symmetry coefficient becomes now mass dependent,
$a_A(A)= a_A^V/(1+ a_A^V/(a_A^S \, A^{1/3}))$.  The standard formula is recovered when ${a_A^V}/{a_A^S}=0$,
i.e.\ the surface does not accept asymmetry excess, or in the limit of $A \rightarrow \infty$.  Within
the modified formula, the symmetry coefficient weakens at low $A$ \cite{mye74,dan03}.  Whether or not the
coefficient may be replaced by the constant $a_A^V$ depends on the ratio $a_A^V/a_A^S$.
That ratio can be determined from the ratio of surface to volume asymmetry partitioning for
the energy minimum in proportion to the capacitances:
\begin{equation}
\frac{N_S-Z_S}{N_V-Z_V}= \frac{C_S}{C_V}
= \frac{A^{2/3}/a_A^S}{A/a_A^V} =  A^{-1/3} \, \frac{a_A^V}{a_A^S} \, .
\label{eq:NZS}
\end{equation}

\section{Asymmetry Skins}

Establishing the relatively small differences in the distribution of neutrons and protons in nuclei has been,
generally, difficult experimentally.
Probes with different sensitivities to protons and neutrons had been utilized, such as
electrons and protons, negative and positive pions, or protons and neutrons, with different associated
systematic errors.  The results have not been expressed
in terms of the surface excess, but rather in terms of the
difference in the r.m.s.\ radii between neutrons and protons.  The conversion from the excess (\ref{eq:NZS})
to the difference of radii is relatively straightforward \cite{dan03}, if the surface diffuseness is similar
for neutrons and protons.  Another issue to consider theoretically is that,
for heavy nuclei, Coulomb forces compete with
symmetry-energy effects, pushing the proton radius out against neutrons and polarizing the nuclear interior.
That competition is easily taken into account by minimizing the sum of three energies with respect to the
asymmetry:
\begin{equation}\label{eq:E3sum}
  E=E_V+E_S+E_C \, ,
\end{equation}
where
\begin{equation}\label{eq:Ec=}
  E_C= \frac{e^2}{4 \pi  \epsilon_0} \, \frac{1}{R} \left( \frac{3}{5}
\, Z_V^2 + Z_V \, Z_S + \frac{1}{2}\, Z_S^2\right) \, .
\end{equation}
From the energy minimization, an analytic formula for the difference of the radii follows,
\begin{eqnarray}
  \frac{\langle r^2 \rangle_n^{1/2} -
\langle r^2 \rangle_p^{1/2}}
{\langle r^2 \rangle^{1/2}}
&
=
&
\frac{A}{6NZ} \, \frac{N-Z}{1 + A^{1/3} \, { {a_A^S}/{a_A^V}}
}
\nonumber \\ &&
- \frac{a_C}{168 { a_A^V}} \, \frac{A^{5/3}}{N} \,
\frac{\frac{10}{
3} +   A^{1/3}\, { {a_A^S}/{a_A^V}}}{1 +  A^{1/3}\,  {a_A^S}/{a_A^V}}
 \, .
 \label{eq:rdif}
\end{eqnarray}
The first term on the r.h.s.\ represents the effects of symmetry energy only, from Eq.\ (\ref{eq:NZS}),
while the second term represents the Coulomb correction.  It should be mentioned that the impact
of the Coulomb-symmetry energy competition is much weaker onto the net energy than onto the skin size.

\begin{figure}
\centerline{\includegraphics[width=.77\linewidth]{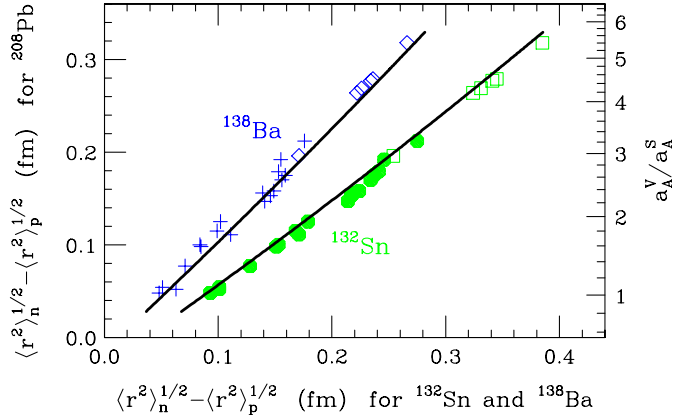}}
\caption{\label{fig:typel}
Correlation between the asymmetry skins for $^{208}$Pb and $^{132}$Sn and $^{138}$Ba,
in the nonrelativistic and relativistic mean-field calculations \cite{typ01} (symbols) and
predicted by macroscopic Eq.~(\protect\ref{eq:rdif}) (lines).}
\end{figure}

Before trying to draw conclusions from data with Eq.~(\ref{eq:rdif}), it may be worthwhile to test
the macroscopic theory against the microscopic.  In their nonrelativistic Hartree-Fock and
relativistic Hartree calculations, Typel and Brown \cite{typ01} observed correlations
between the sizes of asymmetry skins in different nuclei, when utilizing different effective
interactions.  Those correlations are shown in Fig.~\ref{fig:typel} together with the predictions
of Eq.~(\ref{eq:rdif}) when changing the ratio ${a_A^S}/{a_A^V}$.  The accuracy of the macroscopic
theory in reproducing correlations from the microscopic theory appears to be at the level of $0.01$~fm!

\begin{figure}
\centerline{\includegraphics[width=.64\linewidth]{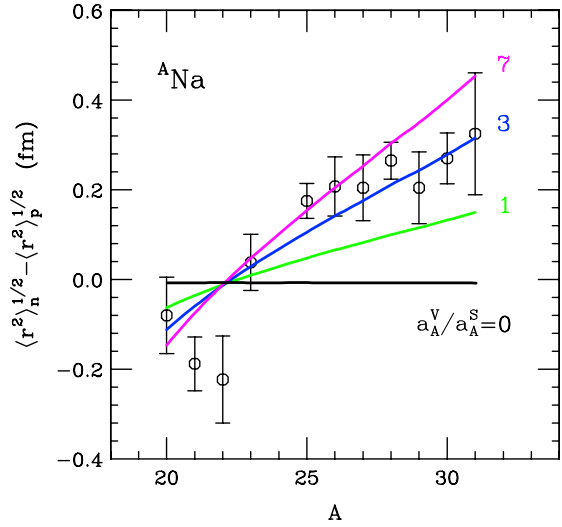}}
\caption{\label{fig:narnp}
Asymmetry skin for Na
isotopes as a function of the mass number, from the data analysis
of Ref.\
\protect\cite{suz95} (symbols) and
from Eq.\ (\protect\ref{eq:rdif}), for the indicated values of
${a_A^S}/{a_A^V}$.}
\end{figure}

\begin{figure}
\centerline{\includegraphics[width=.77\linewidth]{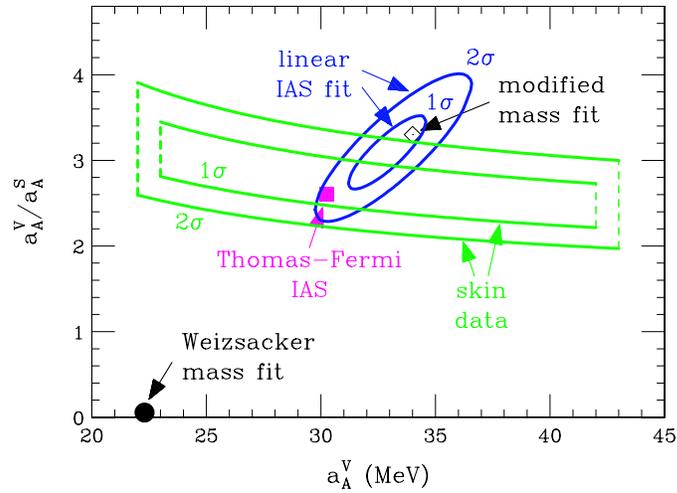}}
\caption{\label{fig:rab}
Constraints on the symmetry energy parameters in the plane of ${a_A^S}/{a_A^V}$
vs ${a_A^V}$.  The sloped horizontal lines represent constraints on ${a_A^S}/{a_A^V}$ from fitting
skin data at an assumed value of ${a_A^V}$.  The elliptical contours represent constraints obtained
from the fit linear in $A^{-1/3}$ to the values of $a_A^{-1}(A)$ from IAS.}
\end{figure}

We next turn to the implications of skin data.  Figure \ref{fig:narnp} shows a comparison of
the data by Suzuki {\em et al.} \cite{suz95} from Na isotopes to the predictions of
Eq.~(\ref{eq:rdif}) for different values of ${a_A^S}/{a_A^V}$.  The comparison suggests a value
of ${a_A^S}/{a_A^V} \sim 3$.  Figure \ref{fig:rab} shows, with sloped parallel lines, the constraints
on ${a_A^S}/{a_A^V}$ from fitting a variety of skin data (for references to the experiments see \cite{dan03}).
Without Coulomb effects the constraint lines would have been horizontal; the weak sensitivity of the fits
to $a_A^V$ results from the second term on the r.h.s.\ of (\ref{eq:rdif}).  The favored values of the
ratio, ${a_A^S}/{a_A^V} \sim 2.8$, imply that $A^{-1/3} \, {a_A^S}/{a_A^V}$ is never small.  Neither can
the $A$-dependent symmetry coefficient be replaced by $a_A^V$, nor even expanded linearly in $A^{-1/3}$.

\section{Isobaric Analogue States}

To find absolute values of the symmetry coefficients, one might try to fit the binding formula to measured
energies.  However, this is treacherous as conclusions on details in different isospin-dependent terms,
including Coulomb, Wigner and pairing get interrelated when drawn from a global fit to the energies.  In
addition, the correlation between mass number and asymmetry along the line of stability correlates the
conclusions on details in the isospin dependent and isospin independent terms.  The conclusions on the
symmetry coefficients change depending on what is done to the other terms in the formula \cite{dan03}.

Optimal for determining the symmetry parameters would be a study of the symmetry term in the binding formula
in isolation from the formula remainder, which might seem impossible.  However, one can take advantage
of the extension of charge symmetry of nuclear interactions to charge invariance.  Under charge invariance
the symmetry term should be a scalar in isospin space \cite{jan03} and can be, thus, generalized with
\begin{eqnarray}
  E_A & = & a_A(A) \, \frac{(N-Z)^2}{A} = 4 \, a_A(A) \,  \frac{T_z^2}{A} \nonumber \\
   & \rightarrow &
  4 \, a_A(A) \, \frac{T^2}{A}= 4 \, a_A(A) \, \frac{T(T+1)}{A} \, ,
\label{eq:EAgeneral}
\end{eqnarray}
where we also happen to absorb most of the Wigner term into the symmetry term.
Under the generalization, the binding formula may be applied to the lowest state of a given isospin $T$ in
a nucleus.  When excited, such a state is an isobaric analogue state (IAS) of the ground state of
a neighboring nucleus.  In the formula generalization, the pairing contribution depends on the evenness of
$T$.

For an isospin of the same evenness as the ground, the change in the formula in the excitation occurs only in the
symmetry term:
\begin{equation}\label{eq:Eexc}
  E_2(T_2)-E_1(T_1) = \frac{4 \, a_A}{A} \big\lbrace T_2(T_2+1) - T_1 (T_1+1) \big\rbrace \, ,
\end{equation}
and the excitation energy can be used to the determine the symmetry energy nucleus by nucleus from
\begin{equation}\label{eq:aAA}
  a_A(A) = \frac{A \, \Delta E}{4 \, \Delta T^2} \, .
\end{equation}
In the context of the previous considerations, the question to ask is whether the deduced $A$-dependent
symmetry coefficient weakens for light nuclei and whether the inverse of the coefficient is linear
in $A^{-1/3}$:
\begin{equation}\label{eq:ainv}
  a_A^{-1} (A) \stackrel{?}{=} (a_A^V)^{-1} + (a_A^S)^{-1} \, A^{-1/3} \, .
\end{equation}

\begin{figure}
\centerline{\includegraphics[width=.77\linewidth]{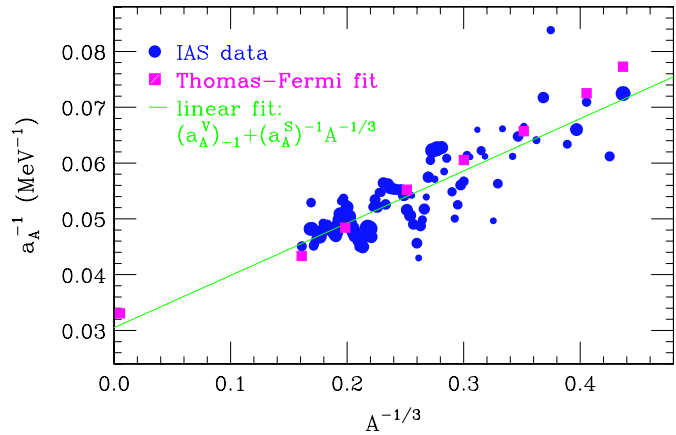}}
\caption{\label{fig:delta}
Inverse of the $A$-dependent symmetry coefficient as a function of $A^{-1/3}$.  Circles
represent values extracted with (\protect\ref{eq:Eexc})
from extremal IAS excitation energies in \protect\cite{ant97}.
Circle size is proportional to the factor $\Delta T^2/A$ in the coefficient
determination.  The line and squares show results of the fits to the experimental results following either
Eq.~(\protect\ref{eq:ainv}) or Thomas-Fermi theory \protect\cite{dan04}.
}
\end{figure}

Inverse values of symmetry coefficients, extracted according to Eq.~(\ref{eq:Eexc}) from IAS data \cite{ant97},
are shown in Fig.~\ref{fig:delta}.  It is seen that the inverse coefficient changes with $A^{-1/3}$
in a roughly linear fashion, although significant shell effects are present.  The line across the figure
represents best fit with Eq.~(\ref{eq:ainv}).  The 1- and 2-$\sigma$ constraints on the symmetry
coefficients from the fit are further indicated with elliptical contours in Fig.~\ref{fig:rab}.
Combining the constraints from the fits, we conclude that
$30.0 \, \mbox{MeV} \lesssim a_A^V \lesssim 32.5 \, \mbox{MeV}$
and
$2.6 \lesssim a_A^V/a_A^S \lesssim 3.0$.

\section{Consequences and Conclusion}

The emergence of the surface capacitance for asymmetry may be tied to the weakening of the symmetry energy
with density, see e.g.\ \cite{bod58,dan03}.
Due to the weakening, it becomes advantageous for the nucleus to push its asymmetry to the
surface to lower energy.  The ratio of the symmetry coefficients, specifically, can be tied to the
shape of the symmetry energy dependence on density as, in the local-density approximation to the symmetry
energy, the ratio is found to be
\begin{equation}\label{eq:aVS}
\frac{a_A^V}{a_A^S} = \frac{3}{r_0} \int dr \, \frac{\rho(r)}{\rho_0}
\, \left[\frac{S(\rho_0)}{S(\rho(r))} - 1 \right] \, .
\end{equation}
Here, the integration is across the nuclear surface and $\rho(r)$ is
the density as a function of position.
For density-independent symmetry energy, $S(\rho)= S(\rho_0) \equiv a_A^V$, the surface does not accept
the asymmetry, ${a_A^V}/{a_A^S}=0$!

Using the correlations between the coefficient ratio, skins and drop of the symmetry energy with $\rho$,
within the relativistic and nonrelativistic calculations by Fuhrnstahl \cite{fur02}, one can arrive at
limits at on the drop, either expressed in terms of the value of symmetry energy at half of the
normal density or in terms of the power of density in parameterization of the symmetry energy \cite{dan04}.
Specifically, one finds
$0.58 \lesssim S(\rho_0/2)/a_A^V \lesssim 0.69$ and $0.54 \lesssim \gamma \lesssim 0.77$ in
$S(\rho) \simeq a_A^V (\rho/\rho_0)^\gamma$.  These further imply limits on pressure in neutron
matter at normal density and, with results of Ref.~\cite{lat01}, produce limits on neutron-star
radius of $11.5 \, \mbox{km} \lesssim R \lesssim 13.5 \, \mbox{km}$ for
$1.4 \, M_{\astrosun}$ mass.


To conclude, the requirement of macroscopic consistency brings in the surface symmetry energy into the nuclear
binding formula.  The volume and surface symmetry energies combine as energies of coupled
capacitors.  The extension of the binding formula implies emergence of the asymmetry skins for nuclei
and weakening of the symmetry term in light nuclei.  The systematic of the asymmetry skins restricts
ratio of the symmetry coefficients.  The charge invariance allows to study variation of the symmetry coefficient
nucleus by nucleus.  Combination of the fits to skins and IAS yields
$30.0 \, \mbox{MeV} \lesssim a_A^V \lesssim 32.5 \, \mbox{MeV}$
and
$2.6 \lesssim a_A^V/a_A^S \lesssim 3.0$.  The surface symmetry energy is associated with weakening of the
symmetry energy with density.  The $a_A^V/a_A^S$ ratio implies limits on drop characteristics, such as
$0.58 \lesssim S(\rho_0/2)/a_A^V \lesssim 0.69$.  Implications for neutron stars follow.
Current direction is to incorporate shell corrections into the IAS analysis.

\end{document}